%% file: aisec2017.tex
\documentclass[sigconf]{acmart}

\usepackage{booktabs} 
\usepackage{url}
\usepackage{fancyhdr}
\usepackage{balance}

\setcopyright{rightsretained}

\acmConference{AISec'17}{November 3, 2017}{Dallas, TX, USA}
\acmPrice{15.00}
\acmDOI{10.1145/3128572.3140445}
\acmYear{2017}
\copyrightyear{2017}

\pdfoutput=1

\begin{document}
\fancyhead{} 
\title{Practical Machine Learning for Cloud Intrusion Detection}
\subtitle{Challenges and the Way Forward}

\author{Ram Shankar Siva Kumar}
\affiliation{%
  \institution{Microsoft}
}
\email{Ram.Shankar@microsoft.com}

\author{Andrew Wicker}
\affiliation{%
  \institution{Microsoft}
}
\email{Andrew.Wicker@microsoft.com}

\author{Matt Swann}
\affiliation{%
  \institution{Microsoft}
}
\email{MSwann@microsoft.com}

\begin{abstract}
Operationalizing machine learning based security detections is extremely challenging, especially in a continuously evolving cloud environment.  Conventional anomaly detection does not produce satisfactory results for analysts that are investigating security incidents in the cloud.  Model evaluation alone presents its own set of problems due to a lack of benchmark datasets.  When deploying these detections, we must deal with model compliance, localization, and data silo issues, among many others. We pose the problem of ``attack disruption'' as a way forward in the security data science space. In this paper, we describe the framework, challenges, and open questions surrounding the successful operationalization of machine learning based security detections in a cloud environment and provide some insights on how we have addressed them. 
\end{abstract}

%


\maketitle

\input{aisec2017_body}

\sloppy
\bibliographystyle{ACM-Reference-Format}
\balance
\bibliography{aisec2017_bibliography} 

\end{document}

%% file: aisec2017_body.tex
\section{Introduction}
The increasing prevalence of cybersecurity attacks has created an imperative for companies to invest in effective tools and techniques for detecting such attacks.  Intrusion detection systems are expected \cite{technavio} to grow to USD 5.93 billion by 2021 at a compound annual growth rate of 12\%.

Academia \cite{tsai2009intrusion,laskov2005learning,eskin2002geometric} and industry have long focused on building security detection systems (shortened hereafter as detection) for traditional, static, on-premise networks (also called ``bare metal'') while research in employing machine learning for cloud setting is more nascent \cite{modi2013survey,roschke2009intrusion,shelke2012intrusion}. Whether detection systems for bare metal or for the cloud, the emphasis is almost always on the algorithmic machinery. This paper takes a different approach - instead of detailing a single algorithm or technique that may or may not be applicable depending on factors like volume of data, velocity of operation (batch, near real time, real time), and availability of labels, we document the challenges and open questions in building machine learning based detection systems for the cloud.  In this spirit, this paper is more closely related to \cite{sommer2010outside} but very specific to building monitoring systems for the cloud's backend infrastructure. 

We report the lessons learned in securing Microsoft Azure which depends on more than 300 different backend infrastructure services to ensure correct functionality. These 300+ services support all flavors of cloud offerings: public cloud (accessible by all customers) and private cloud (implementation of cloud technology within an organization). Within these cloud offerings, the backend services also support different customer needs like Infrastructure as a Service (IaaS) and Platform as a Service (PaaS). Azure backend infrastructure generates more than tens of petabytes of log data per year which has a direct impact on building machine learning based intrusion detection systems. In this setting, seemingly simple tasks such as detecting login anomalies can be difficult when one has to wrestle with 450 billion login events yearly. 

There are other problems besides scalability. Firstly, the cloud environment is constantly shifting: virtual machines are constantly deployed and decommissioned based on demand and usage; developers continuously push out code to support new features which inherently changes the data distributions, and the assumptions made during model building. Secondly, each backend service functions differently. For instance, the backend service that orchestrates Azure's storage solution is architected differently from the backend service that allocates computation power. Hence, to continue with the login anomaly example, one must account for different architectures, data distributions and analyze each service separately. Furthermore, the cloud, unlike traditional systems, is geo-distributed. For instance, Azure has 36 data centers across the world, including China, Europe, and Americas, and hence must respect the privacy and compliance laws in the individual regions. This poses novel challenges in operationalizing security data science solutions. For instance, compliance restrictions that dictate data cannot be exported from specific geographic locations (a security constraint) have a downstream effect on model design, deployment, evaluation, and management strategies (a data science constraint). 

This paper focuses on the practical hurdles in building machine learning systems for intrusion detection systems in a cloud environment for securing the backend infrastructure as opposed to offering frontend security solutions to external customers. Hence, the alerts produced by the detection systems discussed in this paper are consumed by in-house, Microsoft security analysts as opposed to paying customers who buy Azure services. Though not discussed in this paper, we would like to highlight that the frontend monitoring solutions built for external customers are considerably different from backend solutions, as the threat landscape differs based on the customer's cloud offering selection. For instance, if a customer chooses IaaS, important security tasks such as firewall configuration, patching, and management is the customer's responsibility as opposed to PaaS, where most of the security tasks are the cloud provider's responsibility. In practice, the difference between PaaS and IaaS dictates different security monitoring solutions. 

This paper is not about fraud, malware, spam, specific algorithms or techniques.  Instead, we share several open questions related to model compliance, generating attack data for model training, siloed detections, and automation for attack disruption - all in the context of monitoring internal cloud infrastructure. 

We begin with a discussion about building models (or systems) that distinguish between statistical anomalies and security-interesting events using domain knowledge.  This is followed by a discussion of techniques for evaluating security detections.  We then describe issues surrounding model deployment, such as privacy and localization, and present some approaches to address these issues.  We move on to discuss issues with siloed data and models.  We conclude with some ways to move from attack detection to attack disruption.

\section{Evolution to Security Interesting Alerts}
Here is a typical industry scenario: An organization invests in log collection and monitoring systems, then hires data scientists to build advanced security detections only to find that the team of security analysts are unhappy with the results. Disgruntled analysts are not the only thing at stake here: a recent study by the Ponemon Institute \cite{ponemon}, showed that organizations spend, on average, nearly 21,000 hours each year analyzing false positive security alerts, wasting roughly \$1.3 million yearly. To address this issue, it can be appealing to invest in a more complex algorithm that presumably can reduce the false positive rate and surface better anomalies. However, as we describe below, blind adherence to this strategy tends not to yield the desired results. 

As mentioned earlier, Azure has more than hundreds of backend services that are all architected differently. On the one hand, it is impossible to have a single generic anomaly detection that captures the nuances of each service. On the other hand, it is cumbersome to build bespoke machine learning detections for each service. In this section, we describe strategies to combine the regular anomaly detection setting with domain knowledge from the service and security experts in the form of rules to lower false positive rates.

We have established the following criteria for security alerts to help maximize their usefulness to security analysts: Explainable, Credible, and Actionable.  Unfortunately, anomaly detection in an industry setting rarely satisfies these criteria. This is because anomalous events are present in any organization, but not all of these anomalies are ``security interesting'' which is what the security analysts care about.  

As an example, we encountered the following issue when building an anomalous executable detection.  We collaborated with our security investigation team to better understand how attackers masquerade their tools to match common executables.  For instance, attackers would name their tool ``ccalc.exe'' to be deceptively similar to the Microsoft Windows Calculator program ``calc.exe''.  We sought to develop an anomaly detection for finding abnormal executables based on the executable name and metadata.

When we ran this new detection, security experts found most of the alerts were false positives despite conforming to their definition of attacker activity.  For instance, the detection system found an executable named psping.exe that closely resembles ping.exe, but the investigation team found that the service engineers were using a popular system utility tool.  This soon became a recurring theme: the alert appeared worthy of investigation at first glance, but after spending considerable resources on the investigation, we would conclude that the alert was a false positive.

In order to generate useful results, we moved away from simply anomaly detection and focused our efforts on systems that produce ``security interesting'' alerts.  We define such a system as one that captures an adversary's tools, tactics and procedures from the gathered event data while ignoring expected activity. We show later in the section, how rules and domain knowledge can help in these aspects.

\begin{figure}
\includegraphics[height=180bp, width=240bp]{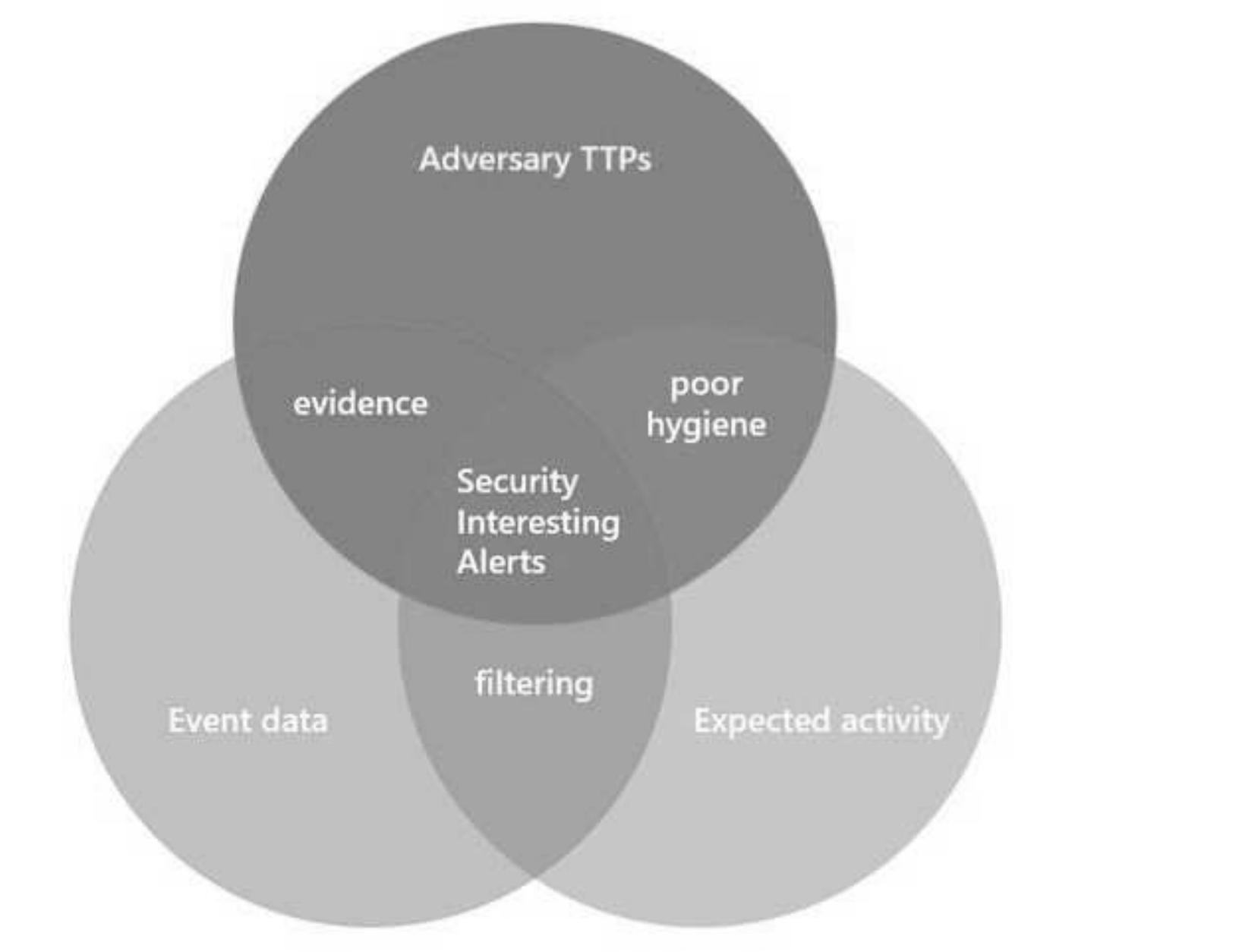}
\caption{A Venn diagram depicting the intersection of security interesting alerts}
\label{evolution_to_security_interesting}
\end{figure}

As a first step, we recommend that machine learning engineers consult with security domain experts to see if there is any overlap between the attacker activity that we seek to detect and expected activity.  If there is some overlap, then this is a ``hygiene'' issue and must be addressed.  For instance, attackers often elevate privileges using ``Run as Administrator'' functionality when compromising infrastructure machines, which can be tracked easily in security event logs.  It is standard operating procedure that service engineers must never elevate to admin privileges without requesting elevated privileges through a just-in-time access system.  This way, the service engineer's high privileged activity is monitored and more importantly, is scoped for a short period of time. However, service engineers often disregard this rule when they are debugging.  This creates a problem in which regular service engineer activity is almost indistinguishable from attacker activity which we refer to this as ``poor hygiene'' (see Figure \ref{evolution_to_security_interesting}).  Specifying and strictly enforcing operating procedures to correct poor hygiene, is the first step in reducing the false positive rate of the system.

Once the hygiene issues are resolved and a well-defined security scenario is in place, the stage is set for incorporating domain knowledge.

\begin{figure}
\includegraphics[height=200bp, width=240bp]{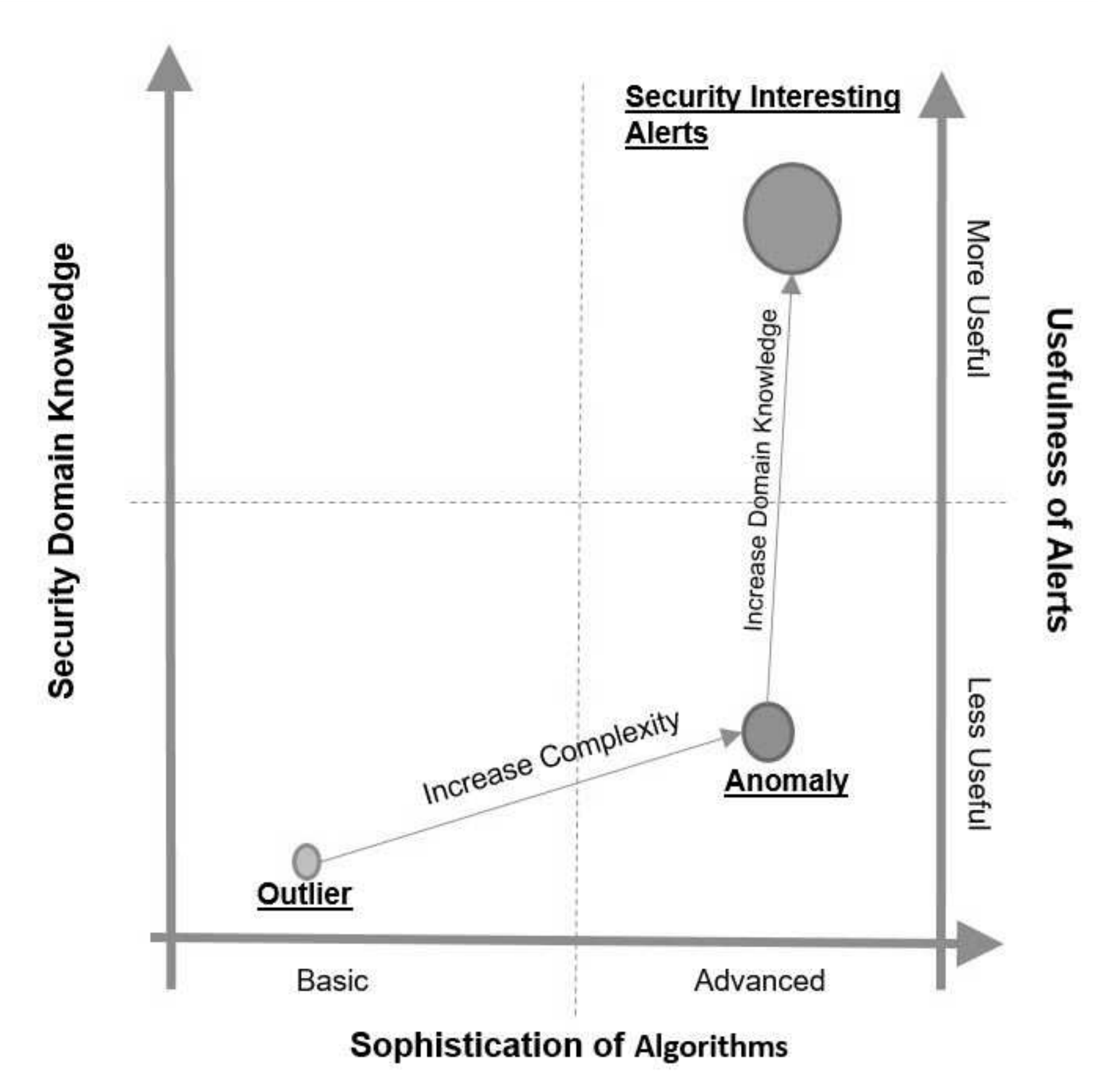}
\caption{Sophistication of anomaly detection techniques}
\label{evolution_to_security_interesting_2}
\end{figure}

\subsection{Strategies to incorporate Domain Knowledge}
Domain knowledge is critical when developing security detections, and how it is leveraged goes well beyond simple feature engineering. In this section, we discuss the different strategies that we have successfully employed to utilize domain knowledge in the form of rules. Other ways to incorporate domain knowledge, not discussed in this paper, are feedback of alerts from security analysts and consuming threat models.  

\subsubsection{Incorporating Rules (end consumer + security experts)}
Rules are an attractive means to incorporate domain knowledge for the following reasons:
\begin{itemize}
	\item They are a direct embodiment of domain knowledge - Most organizations have a corpus of firewall rules (e.g., limiting traffic from Remote Desktop Protocol ports), Web Attack detection rules (e.g., detecting xp\_cmdshell in SQL logs is a strong evidence of compromise), or even direct embodiment of the goodness (like whitelists) and maliciousness (such as blacklists). Security analysts embrace rules because it allows them to easily express their domain knowledge in simple conditionals. If we define rules as an atomic first-order logic statements, then we can expand to a wider set:
	\begin{itemize}
		\item Indicators of Compromise (file hashes, network connections, registry key values, specific user agent strings) that are commonly sourced from commercial vendors; 
		\item Threat intelligence feeds (domain reputation, IP reputation, file reputation, application reputation);
		\item Evidence/telemetry generated by adversary tools, tactics, and procedures that have been observed beforehand.
	\end{itemize}
	\item Rules have the highest precision - Every time a scoped rule fires, it is malicious by construction.
	\item Rules have the highest recall - Whenever a scoped rule fires, it detects all known instances of maliciousness that are observed for that rule. 
\end{itemize}

We also acknowledge the biggest disadvantage of rules: care must be taken to maintain the corpus of rules since stale ones can spike the false positive rate. However, even machine learning models require babysitting and have their own complications \cite{sculley2014machine}. For instance, if we use a model that has been trained on data that no longer reflects the state environment, the model can drift and produce unexpected results. Given that rules encode domain knowledge, are readily available, and favored by security analysts, we present three strategies to incorporate them alongside a machine learning system. 

\paragraph{As filters}
\begin{figure}
\includegraphics[height=60bp, width=240bp]{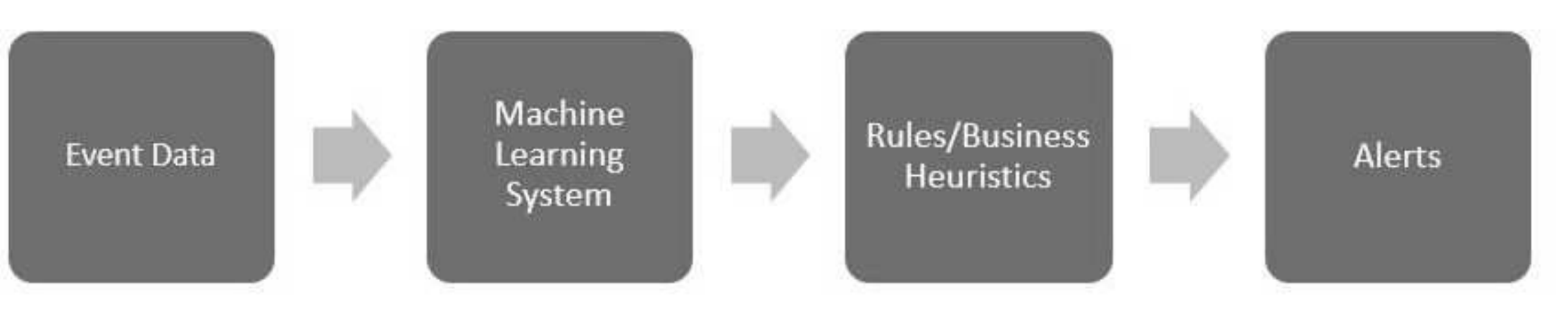}
\caption{Rules can be applied as filters after the machine learning system.  The machine learning system produces anomalies, and the business heuristics help to winnow the security interesting alerts.}
\label{rules_as_filters}
\end{figure}

Rules not only catch known malicious activity, but can also be applied as filters on the output of the machine learning system to sift out the expected activity (see Figure \ref{rules_as_filters}).  In this architecture, the machine learning system produces anomalies, and the rules/business heuristics help to pick out the security interesting alerts.  We used this framework to detect logins from unusual geographic locations.  In this scenario, if a user who always logs in from New York attempts to login from Sydney, then the user must be prompted for multifactor authentication.  Our initial implementation of the detection logic had a false positive rate of 28\%, and at cloud scale, that translated to 280 million ``suspicious'' logins.  To improve our false positive rate, we supplemented the system with custom rules to identify company proxies, cellphone networks, and possible vacations/travel.  After applying such business heuristics, the false positive rate dropped to less than 0.001\% which demonstrate how effective rules can act as filters. 

\paragraph{As binary features}
\begin{figure*}
\includegraphics[height=200bp, width=480bp]{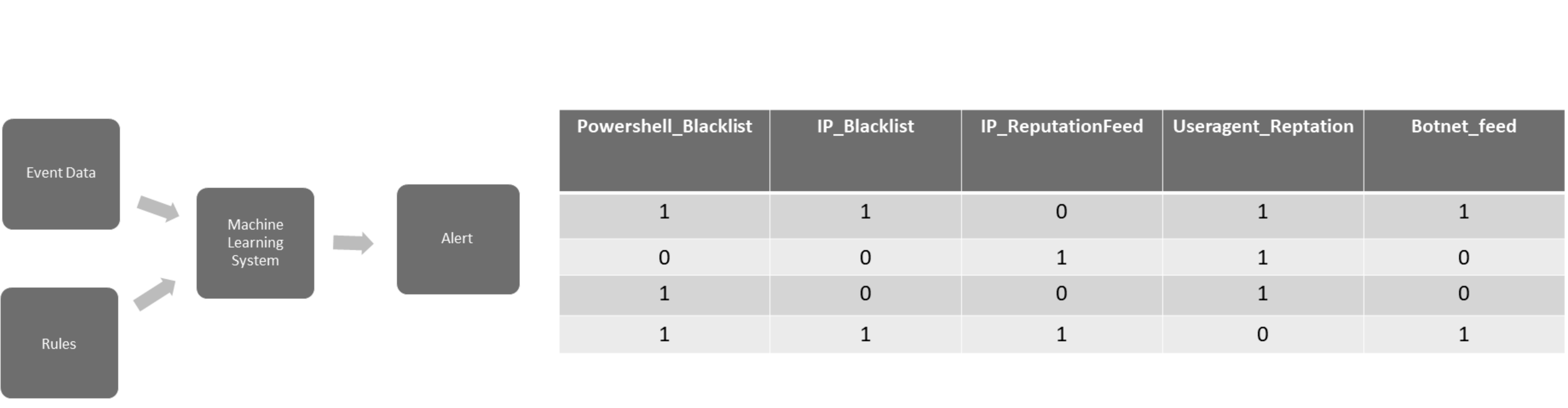}
\caption{Rules can also be used as input to a machine learning system, for example, as binary features.}
\label{rules_as_binary_features}
\end{figure*}

Rules can also be easily incorporated within machine learning systems as binary features (see Figure \ref{rules_as_binary_features}), and is used in our network anomaly detection. For instance, if the rule is TRUE on the event data, it is set to 1, and 0 otherwise. This kind of construction is useful when the rules are explicit and change slowly, such as firewall rules or web attack detection, as opposed to temporary ones (e.g., ``Bob is on vacation from 8/1 to 8/20''). 

\paragraph{Within ML system}
\begin{figure}
\includegraphics[height=160bp, width=240bp]{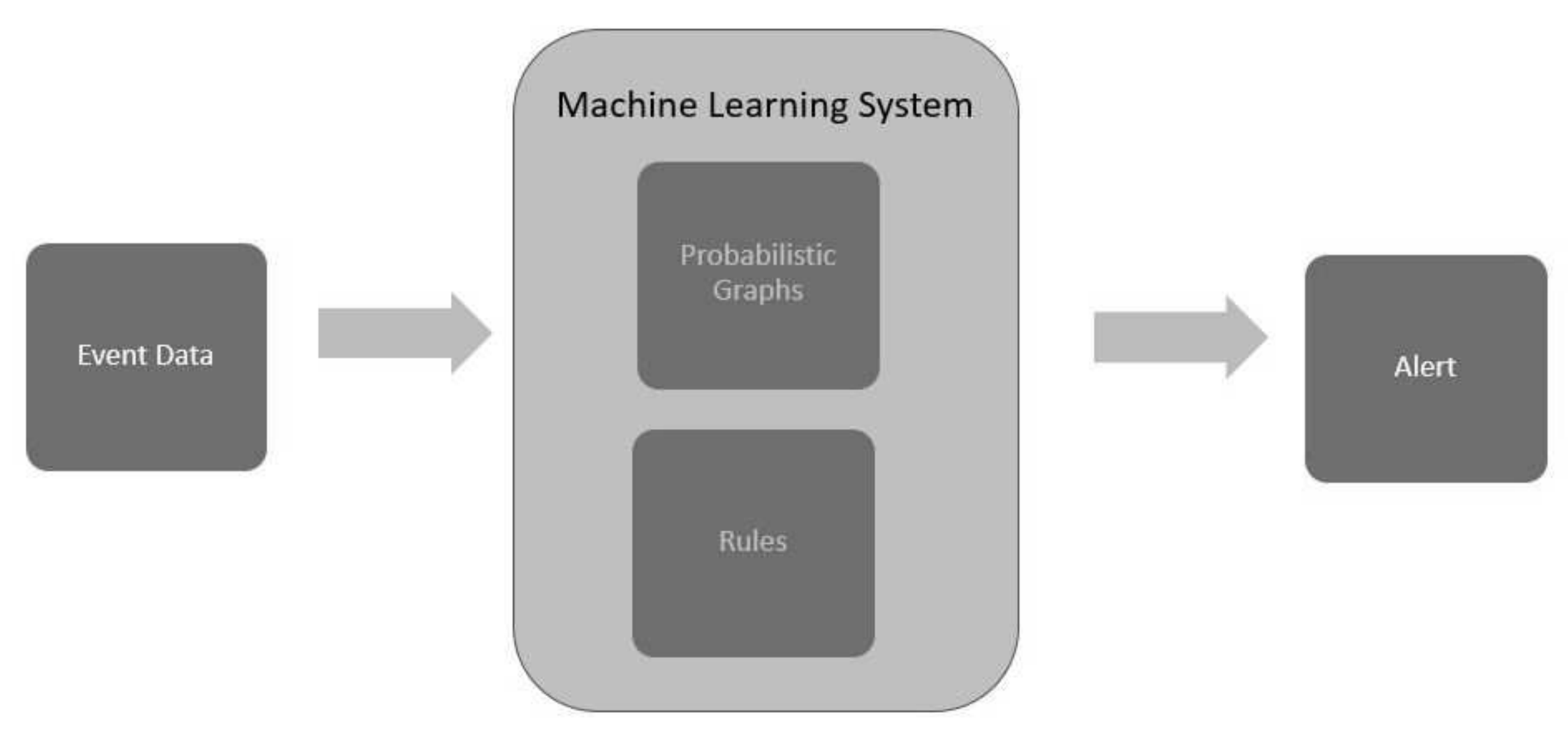}
\caption{Representations such as Markov Logic Nets and Probabilistic Soft Logic help to combine rules and probabilistic representations as one ML entity.}
\label{rules_within_ml_system}
\end{figure}

Unlike the first two constructions where the corpus of rules was separate from the machine learning system, it is also possible to encapsulate the rules and machine learning system as one entity (see Figure \ref{rules_within_ml_system}). The biggest advantage is that when this embodiment is used, there is no need to maintain two separate code bases (code base for machine learning + code base for filters vs. code base for machine learning + code base for rules vs code base for purely machine learning system).  Security scenario formalizations in terms of Markov logic networks \cite{richardson2006markov} and probabilistic soft logic \cite{kimmig2012short} allow us to combine first order logic (in our case rules) and probabilistic graphs. In our experience, the biggest drawbacks with these systems is that the amount of time for inference quickly becomes untenable for large scale cloud services. We leave it as future work to explore the applicability of systems such as Markov logic networks to the security data science community. 

\section{Model Evaluation}
In the race to build successful security analytics, an inordinate amount of attention has been paid to selecting the most appropriate algorithms. However, the challenge is not model building, but convincing end-consumers such as security analysts or on-call service engineers that the system works. This is a rather difficult task in practice, owing to a paucity of labeled data and more importantly, a lack of baseline datasets for the intrusion detection problem in the cloud space.

Consider some of the popular benchmark data sets: ImageNet \cite{imagenet_cvpr09} dataset offers 14 million labeled images for image recognition problems; the SwitchBoard corpus \cite{godfrey1997switchboard} offers close to 5,800 minutes of labelled data for speech recognition problems.  Security applications, on the other hand, have outmoded datasets - the last relevant data set goes back to the 1999 KDD data set \cite{kddcup1999data} on network intrusions from an artificially created network by MIT Lincoln Laboratory. 

However, for the cloud setting there are no prevalent benchmarks, and we argue that such benchmarks may never exist for the following reasons:
\begin{itemize}
	\item Simulated environments with artificially created attacks represent a static, sterile setting that is hardly the case in practice. In a cloud setting, for instance, VMs are continuously being provisioned and decommissioned; deployments happen erratically, and the underlying environment is constantly shifting. 
	\item Attacks in the real-world leave very few traces in most common sources of telemetry (e.g., the windows security event log), which can be as low as two entries in the logs of interest. This is too small a sample of labeled data to measure against.
	\item One benefit of a benchmark data set such as MNIST is that we can make a closed-world assumption about the data.  Everything that we need to know about our data is contained entirely within the image itself, with the addition of a simple label.  In a security setting, we can rarely make a closed-world assumption, as with image recognition systems.  For example, the audit logs from an authentication service might not tell the complete story about suspicious activities it, in-part, captures.
	\item Unlike images where the underlying data is always pixels, log data is extremely specific to the organization. While there is some standardization such as NetFlow, the logs from different environments tend to have few similarities, and more importantly have different assumptions. Therefore, training or testing on another company's logs would be less effective.
	\item Unless it is a ``threat intelligence'' company selling scoped tainted data as indicators of compromise, there is no incentive for a compromised cloud company to share their raw logs of tainted data to the public, as it may invite further scrutiny. 
\end{itemize}

For these reasons, security data scientists must create their own baseline datasets. In this section, we will discuss the different strategies that we use in practice to evaluate detection systems. There are two steps: 
\begin{itemize}
	\item Use security knowledge to bootstrap the intrusion detection system to provide a starting point of attack data
	\item Use machine learning techniques to grow this seed of labeled data
\end{itemize}

\subsection{Bootstrapping using security strategies}
For the bootstrap step, we present three strategies with each strategy mapping to different maturity levels of the detection lifecycle.  For instance, at the beginning stages of developing detection systems, we can run a quick test by injecting synthetic data to check if the system is operational. Once the system has become acceptably mature, we can use a red team. 

\subsubsection{Trivial case: Inject fake malicious data into data storage}
The idea here is to create fabricated attacker activity, and test it against the detection system. We can validate that the system works as intended if it can surface the injected data among the rest of the non-anomalous data. While this may seem trivial, we regularly use it as a test to see if the system is functioning. For instance, when the Office 365 security team built a Markov chain model (MCM) to detect unusual operating system activity, they had no test data to verify if the system worked as intended. So, they simply created an account, ``eviluser'', and associated it with arbitrary activity. They knew their system was functioning as intended if ``eviluser'' would rank on top of the alert stack. In some ways, this became a poor man's debugger. Of course, with such low overhead comes a huge tradeoff: the injected data may not be representative of true attacker activity. It is important to note that we do not espouse validating or generating evaluation data from seemingly random data. Instead, we are attesting how synthetic data can act as sanity checks to test detection systems.  

\subsubsection{Cross Pollinate from security products}
A more advanced strategy for gathering evaluation data is to utilize existing security products and attacker tools. Here are some strategies we employ:
\begin{itemize}
	\item Use common attack tools: If the task at hand is to detect malicious processes running in the OS, engineers can run tools like Metasploit \cite{metasploit}, PowerSploit \cite{powersploit}, or Veil \cite{veil}in the environment, and look for traces in the logs. This technique is similar to using fuzzers for web detection. While easy to implement, care must be taken not to modify the model based on this data. This way, we avoid overfitting to the tool instead of generalizing to real-world attacks. 
	\item Data from existing security products: If an organization has parallel security products, they all are potential for new label data. For instance, when we wanted to test our detection for compromised VMs in our cloud infrastructure, we leveraged our online mail service for labels of compromised data. Specifically, we used our mail service which indicated the IP addresses that sent spam. For those IP addresses that stem from our own cloud service, we were able to track down the VMs. Since these VMs were sending out spam, we assumed they were compromised and used their logs to evaluate our detection. The more diverse an organization's security product line is, the easier it is to leverage data for evaluation. 
	\item Honeypots: If it is not possible to acquire data from existing security products, then honeypots can be deployed.  When attackers ultimately compromise honeypots, logs from these systems can be used for evaluation.  There are two drawbacks to this strategy:
	\begin{itemize}
		\item Not all attacks can be captured via honeypots. For instance, we see a preponderance of network reconnaissance (such as ping, and checking for open SSH ports) as opposed to more arcane attacks such as hypervisor escapes. 
		\item There is no guarantee that attackers will be lured by the honeypot, and hence no guarantee that we will receive any evaluation data. In other words, we will only be able to collect data for opportunistic attacks and not targeted ones.  
	\end{itemize}
\end{itemize}

\subsubsection{Test against red team}
The highest quality evaluation data is generated when we hold penetration test engagements in which authorized engineers play the role of an adversary (commonly called ``red team'') and attempt to accomplish their goals while subverting our detection system. Red team members emulate adversaries with specific objectives, like exfiltrating data and begin from scratch with reconnaissance. Results from red teaming are the best approximations to how our systems are attacked ``in the wild'' and make a great way to validate detection systems. There are some caveats to using red teams: firstly, red teams are expensive resources, and in most cases, only large organizations have in-house red teams. Secondly, it is important to note that red team engagements are point-in-time exercises that are typically scheduled once every four months whereas security analyst teams are constantly building new detections based on the currents threats. Finally, evaluating the detection system after a red team engagement is not trivial. For instance, even when red team members take meticulous notes during the process, identifying the malicious sessions is an extremely time-consuming task. For example, their logs might read ``ran tool X'' with a timestamp. However, security analysts would need to spend an inordinate amount of time mapping back to identify the attack data in the logs. 

\subsection{Machine learning for increasing attack data}
It is important to note that data produced by the red team or cross-pollination is not sufficient by itself for model evaluation. For instance, the data generated by red team activities is limited because the engagement is scoped: red teams tend to compromise only a handful of accounts (in the order of tens as opposed to thousands) that are required to achieve their objective. These precise movements leave very few traces in logs. To tackle this problem, in this section, we describe how to employ machine learning techniques to increase our attack data corpus. 

It is not good strategy in practice to undersample the majority class (i.e., the ``normal'', non-malicious data) as it is tantamount to throwing away data - the model never sees some of the normal data, which may encode important information. On the other hand, naively oversampling the minority classes (i.e., the generated attack data) has the disadvantage of repeating the same observed variations - we are not increasing the data set; we are merely replicating it. Instead, we recommend SMOTE \cite{chawla2002smote}.  Essentially, it generates a random point along the line segment connecting an anomalous point, or ``attack point'', to its nearest neighbor. In our experience, SMOTE outperforms weighted decision trees. 

Recent developments in Generative Adversarial Networks (GAN) \cite{goodfellow2014generative} suggest they are a promising vehicle for increasing attack corpus. Results from \cite{hu2017generating,grosse2016adversarial} show that GANs can produce adversarial examples that successfully trick the intrusion detection system.  We posit that these same techniques can be used to synthesize attack data for evaluation. There are only two perceived drawbacks: the bootstrap dataset must be sufficiently large (while this may be easy in the case of malware analysis, it is going to be difficult to get many examples of insider attacks for intrusion detection).  We have also found that GANs are particularly tricky to train. 

\section{Model Explainability}
For any practical application of machine learning to security, there is no perfect model.  We must accept that security signals will be wrong sometimes; security analysts are well-aware of this fact.  In a cloud setting where there can be thousands of alerts, model explainability becomes a major issue. For example, it is not sufficient to say there was an unusual .exe file that executed on the system. The alert must convey why the .exe file was anomalous - perhaps it ran from tmp folder; perhaps it took base-64 encoded inputs (which attackers always do); perhaps the .exe file stemmed from a browser, right after downloading a .pdf document. Giving security analysts such explanations is instrumental in eliciting actions. 

When creating new security signals, we must consider that the alerts generated by the signal should be explainable.  The earlier this is taken into account during the signal development process, the better.  Many common and otherwise useful techniques do not easily facilitate explaining the results they produce. 

For example, Deep Learning produces good results in many applications, but it can be difficult to explain why the model produced the results that it did.  On the other hand, the output of a simple linear model can often be explained in terms of the coefficients. 

The explanations provided can come in different forms.  The most obvious, at least from a machine learning perspective, is to provide one or more features of the model that most impacted the score.  That is, we can provide the features that ``caused'' the model to score the event as suspicious.  This does not need to be a statistically sound explanation.  Interaction effects among features can cause us to provide many features as explanations, but this only creates more information overload for the analyst.  Instead, we should treat the explanation as breadcrumbs that can lead the analyst on the right path.

Some other ways in which we can help explain the results of a security signal are to provide supplemental data to the output and to provide a textual description.  Supplemental data can be something as simple as a ranked order of the suspicious executables on a machine, that can possibly help explain a signal's output or, at least, put it in a context that helps the analyst make sense of the output.  A textual description is not always straight-forward, but simplifies the need to further interpret the explanation.  For example, a suspicious login signal might provide an explanation such as ``high speed of travel to an unfamiliar location''.  This can be much more meaningful that having the analyst look at a set of real-valued features to draw the same conclusion.

Although the output produced by some machine learning models is not easily explained, there have nonetheless been some attempts at providing some explanations.  Among these are techniques such as Local Interpretable Model-Agnostic Explanations (LIME) \cite{ribeiro2016should}.

\section{Model Compliance and Localization}
Most cloud services have data centers across the world - Azure, for instance, has 36 data centers spread across every continent except Antarctica. In this section, we explore how compliance laws across the world affect building security data science solutions space along with a closely related problem of localization.

Building models that are respectful of compliance laws is important because failure to do so not only brings with it crippling monetary costs - for instance, failure to adhere to the new European General Data Protection Regulation (GDPR) set to take effect on May 2018, can result in a fine up to 20 Million Euros or 4\% of annual global turnover \cite{gdpr} - but also the negative press associated for the business. However, building privacy compliant models presents three challenges:
\begin{itemize}
	\item Data protection laws are not uniform across the world. For instance, in the European Union, Article 29 of the Data Protection Working PARTY \cite{personaldata} unequivocally considers ``IP addresses as data relating to an identifiable person'', whereas in the United States, IP address by itself is not personally identifiable and United States courts have opined that ``IP address identifies a computer not a person'' \cite{johnson2009}. Since we want to be able to glean as much information as possible, we would need to follow different masking procedures in different regions which complicates model building and model deployment.  
	\item Data protection laws ask for retroactive modification. To use GDPR as a running example, Article 17 requires right to erasure or commonly called ``Right to be forgotten'', wherein companies must delete records pertaining to a person, when asked. This brings a lot of challenges to the security setting. Consider the simple case of building a system that computes user risk based on user activity. One of the common techniques to evaluate risk score is ``guilty by association'' wherein a user's risk score is increased if the user communicates with other compromised or suspicious users. In other words, user risk score is modeled jointly with other users and not in isolation.  Now, because of right to be forgotten, if a user's details are deleted, should the user's risk score that was used as input to calculate other risk score also change? How do we make models elastic to retroactively delete results and yet remain logically consistent? This would be a good question to explore for future research. 
	\item Privacy laws change with the political landscape - for instance, it is not clear how Britain leaving the European Union would affect the privacy laws in the country. 
\end{itemize}

Closely related to model compliance is model localization. For instance, we built a login anomaly detection using seasonality to detect anomalous login during ``off hours'' (common wisdom stating that anyone logging at midnight on a weekend to access confidential material, is anomalous). However, this na\"ive system quickly backfired - timestamps were normalized to GMT standard, and converting to local time was not possible because there was no IP address, owing to privacy reasons, for reverse geo mapping. Also, in a global company, the definition of a weekend is labile - weekend in the Middle East is not the same as weekend in the Americas. Another complication was that because product adoption happens at different rates across the world, the data generated by each region is different. This is because the data distribution from a mature market is very different that of an emerging market, and hence dictates different models.

\begin{figure}
\includegraphics[height=170bp, width=240bp]{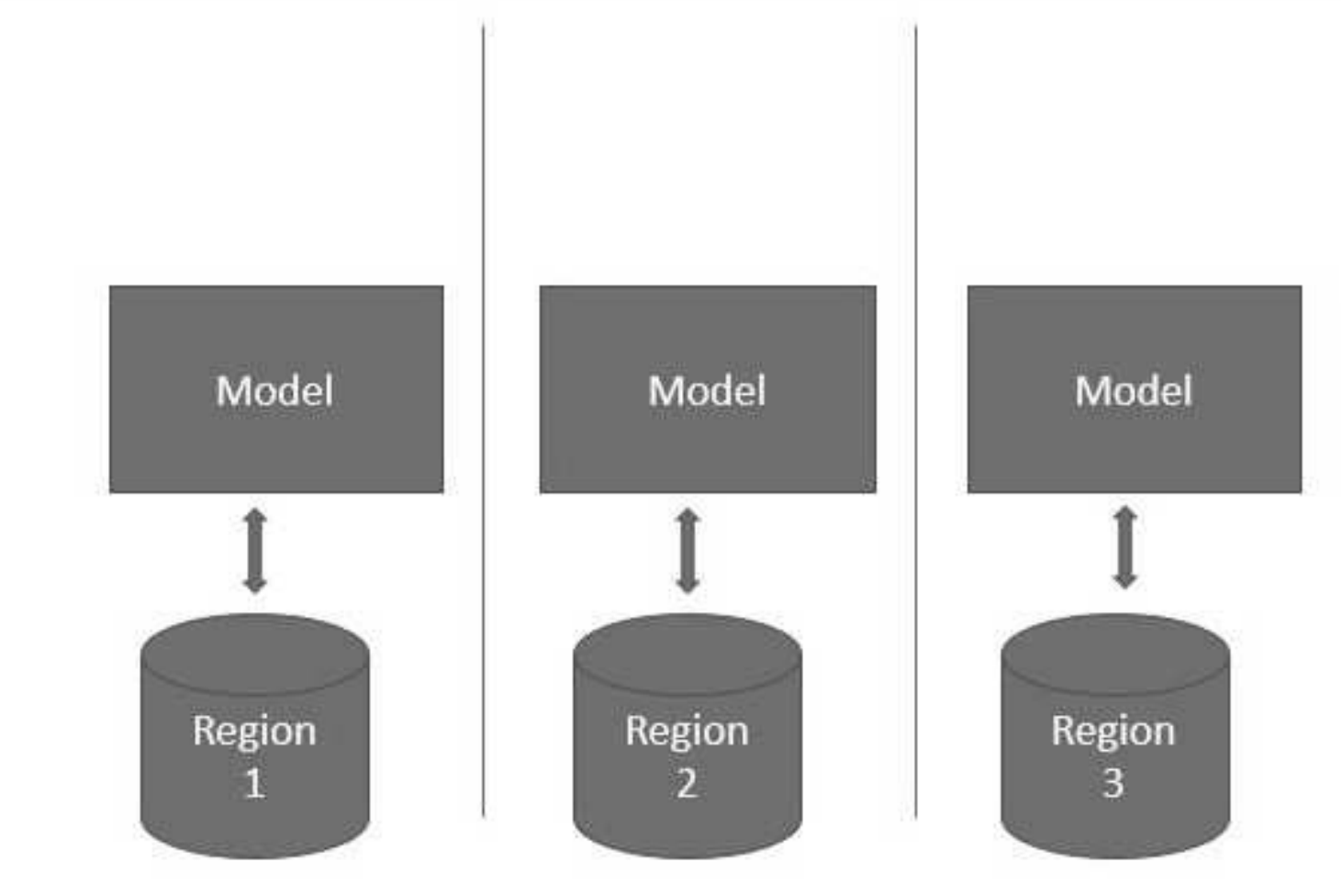}
\caption{Shotgun deployment places same model code across different regions. There are two disadvantages: The model may not be appropriate for a particular region. Also, there is no communication between models.}
\label{model_compliance_1}
\end{figure}

Should there be no model compliance and model localization problems, one can simply use ``shotgun deployment'', i.e., deploy the same model code across different regions (see Figure \ref{model_compliance_1}). With this comes deployment ease as there is no customization and from an operational and debugging perspective, our support staff are happy because there was only one troubleshooting guide to debug any failure. However, because the same model template is used, we are forced to make the erroneous that the data distribution is same across the world.  

\begin{figure}
\includegraphics[height=220bp, width=240bp]{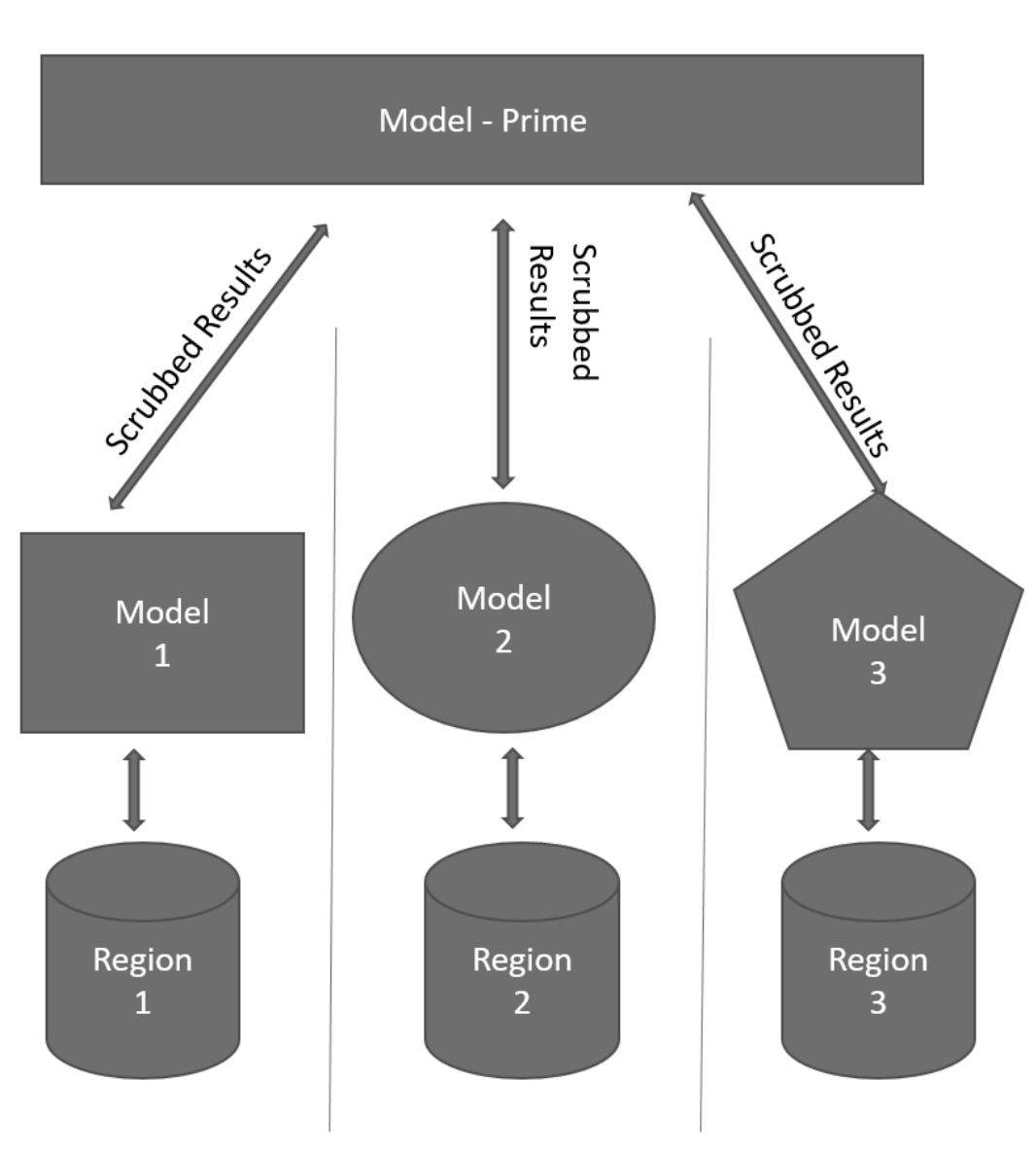}
\caption{Tiered Modeling: Each region is modeled separately. Results are scrubbed according to compliance laws and privacy agreements. Scrubbed results are used as input to ``Model Prime'', where they are collated for global trends.}
\label{model_compliance_2}
\end{figure}

For capturing global trends, we propose ``tiered modeling'' - wherein each geographic region is modeled separately using bespoke models which then send the scrubbed results to a central ``model-prime'' (see Figure \ref{model_compliance_2}). The ``model-prime'' reasons over the output of the individual regions to centralize for global trends. This architecture has two advantages: by having models that run in the context of a region, it is easier to respect local privacy laws and account for nuances of the region; at the same time, having a model-prime accounts for global trends. This comes at a cost: in our experience, this architecture requires specialized deployment frameworks and code maintenance or instance, code change in one region, most likely requires retraining of model-prime. 

This idea of tiered modeling is built on the idea of differential privacy \cite{dwork2008differential} and is similar to the work on applying differential privacy in an ensemble setting \cite{hamm2016learning}, with the main exceptions: tiered modeling is an ensemble of different kinds of learners as opposed to a single type of learner (to ensure that the compliance and localization complexities in different regions, may require different types of learners).  We leave it as future work to study the theoretical guarantees of tiered modeling.  

\section{Horizontally and Vertically Siloed Approaches}
In addition to the problems surrounding anomaly detection, model evaluation, and compliance, we are also faced with complicating factors derived from data silos.  These data silos make it difficult to build security detections across various data sets due to access boundary restrictions.  In this section, we describe two types of data siloes, referred to here as horizontal and vertical siloes, and present some of the problems encountered with each type.  

Access to data required to create and run new security signals is often restricted due to privacy concerns, among other reasons.  This leads to horizontally siloed detections.  In a system like Azure where there are hundreds of different services, attacks can span multiple services.  For example, an attacker might first compromise the service responsible for authentication and then proceed to the service that controls storage. In this case, it is ineffective to build security signals that might protect a single service but have no visibility into other services. 

There has been some success applying signal fusion techniques to the problem of horizontally siloed signals.  We can take various signals from different services that are otherwise siloed, then apply a signal fusion algorithm to combine their outputs.  For example, the outputs of many signals can form features in a simple linear model, or we can leverage ensemble machine learning to combine the individual signals (see \cite{dietterich2000ensemblemethods} for an overview of such techniques).

Vertically siloed detections present a unique set of challenges.  When creating a new security detection, we must decide what is the appropriate level of abstraction that is needed to reliably detect the security events of interest.  As an example, determining if a process is being exploited for an in-memory malware requires a low-level of abstraction since we must look for suspicious memory allocation and suspicious DLL loading.  On the other hand, detecting unusual logins for accounts generally requires a higher-level of abstraction so we can monitor suspicious login locations and suspicious ``impossible travel'' logins.  These types of abstractions are often necessary and sufficient for the intended scenarios.  However, the different levels of abstractions can obfuscate event data that can differentiate a benign event from a malicious event.  

Consider a sequence of actions taken on a cloud storage service.  We might have events with actions such as FileOpened, FileUploaded, and FileDeleted.  These events might contain supplemental information such as UserAgent and Timestamp, but there is only so much that can be used to differentiate a benign action from a malicious one at this level of abstraction.  The lower level actions associated with this event can help inform our judgment.  For example, a password-protected malicious document requires user input to open, so our higher-level abstraction signals might view this as benign.  In this case, our lower-level signals can inform the higher-level signals, even if none of the signals can individually conclude with high confidence the exact nature of the event.

This issue of vertically siloed detections is not simply a consequence of data access boundaries.  It is primarily a consequence of a lack of ability to reason across these abstraction levels.  Unfortunately, it is deceptively challenging to reason across these abstraction levels.  We cannot simply apply a signal fusion algorithm as with horizontally siloed signals.  This is because the very nature of abstraction means that signals built at a higher level of abstraction are affected by data captured at a lower level of abstraction.  This can create a sort of redundancy in which, essentially, the same thing is being measured more than once.  The effect is that the wrong things can be reinforced while the right things are overlooked.

There are many complicating factors with combining the results of vertically siloed detections, and they are not always obvious in a cloud environment with a large number of services.  Some of these problems have been addressed in related problem settings \cite{sculley2014machine} but there is still opportunity for substantial progress to be made here. 

\section{Way forward: Attack Detection to Attack Disruption}
While we continue to bolster our detection systems for the cloud, we would like to draw the attention of the security data science community to what we think is the next wave of advancements in this field. 

We begin with revisiting the adversary's kill chain \cite{hutchins2011killchain}, which details the most common sequence of steps an attacker follows in an industrial setting. It describes how an attacker first performs reconnaissance, delivers malware (most commonly via phishing) to establish foothold; establishes persistence by installing rootkits. From here on, the adversary moves from machine to machine in search of the goal. Once the attacker finds the goal, he or she elevates to administrator privileges and, in most settings, exfiltrates the data of interest. 

After carefully analyzing several security breaches, and interviewing internal security analysts (the ``blue team''), we discovered that blue teams execute their own steps to evict the adversary from the environment which runs parallel to the adversary kill chain. Specifically, blue team members perform the following steps: 
\begin{itemize}
	\item Detect the evidence as an indication of compromise
	\item Alert the appropriate security team
	\item Triage the alert to determine whether it warrants further investigation
	\item Gather context from the environment to scope the breach
	\item Form a remediation plan to contain or evict the adversary
	\item Execute the remediation plan
\end{itemize}

\begin{figure*}
\includegraphics[height=100bp, width=450bp]{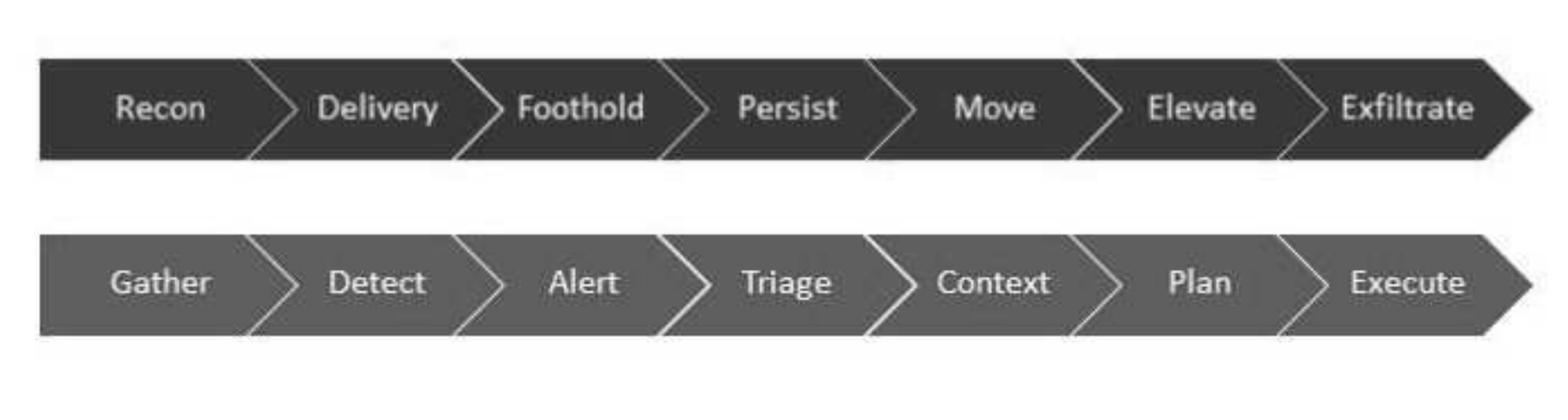}
\caption{Kill chains}
\label{attack_disruption_1}
\end{figure*}

This formulation is pertinent because industry reports \cite{solutions2017data} show that ``compromises are measured in minutes 98\% of the time'' whereas mean time to detect breaches is in the order of months or longer. While we must detect that an attack has happened - even if it's long after exfiltration has finished - the only strategy that is of value to the business is attack disruption. If we were to use the analogy of a building on fire, the blue team needs to transition from the role of an arson investigator and to that of a firefighter. 

\begin{figure}
\includegraphics[height=130bp, width=240bp]{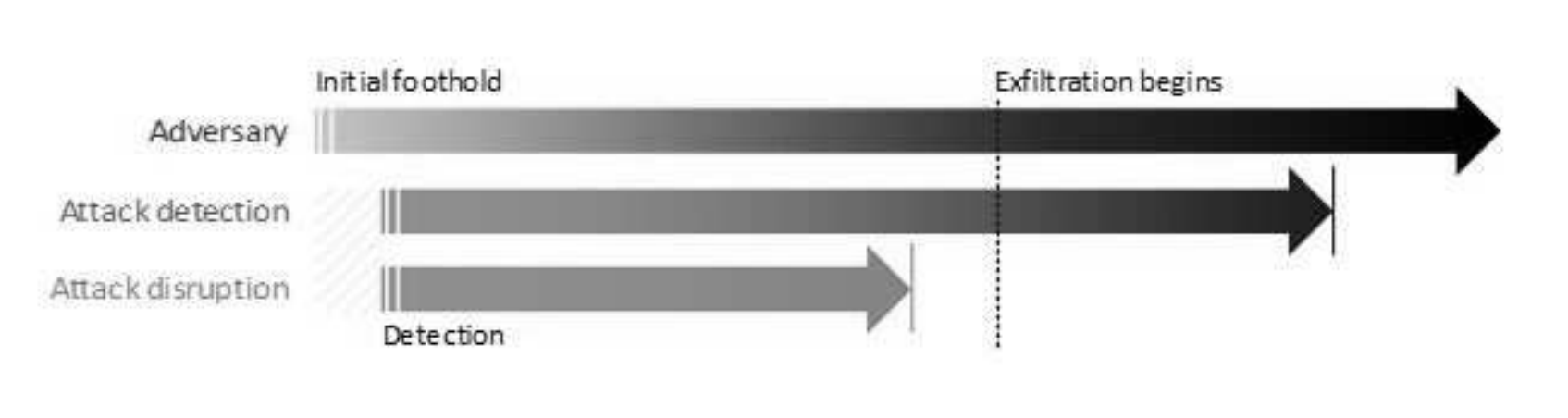}
\caption{Kill chain attack disruption}
\label{attack_disruption_2}
\end{figure}

There is more to attack disruption than a pressing need to act faster via automation (such as ``automatic incident response'', ``automatic remediation'').  Is there a place for intelligence across the blue team kill chain - specifically, is there a place for machine learning to achieve this goal of attack disruption? For instance, can natural language processing and chatbots help analysts in triaging alerts; can recommender engines recommend the next steps in investigations based on previous experiences? Are the techniques used in network traffic optimization transferable for remediation so that traffic from tainted servers is appropriately sinkholed? Applying machine learning to attack disruption has many open questions, and we urge the security data science community to think about this space. 

\section{Conclusion}
In this paper, we described the difficulties in building intrusion detection systems for the cloud. We have claimed that conventional anomaly detection, by itself, does not produce useful alerts in a cloud setting. In practice, we find that a hybrid approach of rules and machine learning yields better results, and showed how they can be combined in the form of filters, features, or even as one single machine learning unit. Since there is no benchmark for evaluating cloud intrusion detection systems, we outlined strategies for gathering high quality evaluation data using other security products or red teams (recommended) and grow the dataset using SMOTE or possibly GANs. We offered our experience in model explainability, and demonstrate how it is more important than ever.  Because of the geo-distributed and global nature of the cloud, we must then deal with model compliance, localization, and data silo issues that affect model design and development. Finally, we shared a framework for attack disruption as the way forward and look to the security data science community for intelligent automation of the blue team kill chain. 

\begin{acks}
We would like to thank Bryan Smith, Eugene Bobukh, Asghar Dehghani, Anisha Mazumder, Haijun Zhai, and Bin Xu for their valuable comments and members of Identity Driven Machine Learning team, Azure Red team, Office 365 Security, Azure Security Center, Cloud + Enterprise Threat Protection team, Applications Security Group and Windows Defender Security for engaged discussions. 
\end{acks}